\documentclass[a4paper]{article}
\usepackage{graphicx}
\usepackage{onecolceurws}
\usepackage{xspace}
\usepackage{url}
\usepackage{verbatim}
\usepackage{bm}
\usepackage[show]{ed}
\def\argmax{\mathop{{\rm argmax}}}
\def\R{{\bf R}}
\def\Z{{\bf Z}}
\newcommand{\scsc}{\textsf{SC$^\mathsf{2}$}\xspace}

\title{OpenMath and SMT-LIB}

\author{
James H. Davenport\\ Department of Computer Science\\
                University of Bath, Bath U.K. \\ J.H.Davenport@bath.ac.uk
\and
Matthew England\\ Faculty of Engineering, Environment \& Computing\\
                Coventry University, Coventry, U.K. \\ Matthew.England@coventry.ac.uk \\ \,
\and
Roberto Sebastiani and Patrick Trentin\\
Dipartimento di Ingegneria e Scienza dell'Informazione (DISI), \\
Universit\`a di Trento, Treno, Italy\\  \{roberto.sebastiani, patrick.trentin\}@unitn.it
}

\institution{}

\begin{document}
\maketitle

\begin{abstract}
OpenMath and SMT-LIB are languages with very different origins, but both ``represent mathematics''.  We describe SMT-LIB for the OpenMath community and consider adaptations for both languages to support the growing \scsc initiative.
\end{abstract}
\vskip 32pt

\section{Motivation: The \scsc Project}

The authors are all members or associates of the EU-funded Horizon 2020 Project \scsc.  The  overall aim of the project is to create a new research community bridging the gap between \textbf{S}atisfiability \textbf{C}hecking and \textbf{S}ymbolic \textbf{C}computation, so that members well informed about both fields can ultimately  resolve problems currently beyond the scope of either.  

The project was motivated by the movement of the Satisfiability Checking community outside of the Boolean SAT problem to consider how their techniques may perform on other domains, creating the field of Satisfiability Module Theories (SMT).  The main idea here is to combine the sophisticated technology built for SAT with calls to domain-specific theory solvers when information beyond the logical structure is required.  Most recently SMT has started to include the domain of non-linear polynomials over the reals, a field of study since the early days of Symbolic Computation.

However, as described in \cite{Abraham2015} it is not sufficient to call leading Computer Algebra Systems as theory solvers.  Rather the algorithms need to be adapted to make them suitable for SMT.  The two communities now find themselves addressing similar problems and so will share the challenge to improve their solutions to achieve applicability on complex large-scale applications. For further details we refer to the project introduction paper at CICM 2016 \cite{Abrahametal2016b} and the project website: \url{http://www.sc-square.org/CSA/welcome.html}.

\subsection{An \scsc Goal: Extending SMT-LIB}

The increasing variety of the theories considered by SMT solvers created an
urgent need for a common input language. The SMT-LIB initiative provided this and a large and increasing number of benchmarks.  Once a problem is formulated in the SMT-LIB language, the user can employ any SMT solver to solve the problem.  It is housed at: \url{htpp://www.smt-lib.org}.  The initiative has proved to spur on research, providing the basis for competitions and collaboration.

Although a proven valuable resource for the general SMT community (and far surpasses anything in Symbolic Computation), SMT-LIB has been found lacking on the domains relevant to \scsc.  There is an NRA (Non-linear Real Arithmetic) category of the SMT-LIB benchmark library with several thousand problems in, but according to \cite{JdM12} this consists mostly of problems originating from attempts to prove termination of term-rewrite systems.  It has been noted in several papers how many of the problems are trivial (solved without calls to theory solvers) or come from a small number of classes and may have some hidden uniformity (see for example \cite{ED16b}).

However, the work needed here is more than a greater variety of benchmark problems.  Rather the depth of problems that can be tackled in this domain requires an extension of the SMT-LIB language instead. Indeed, this is one of the specific \scsc goals, as described in \scsc \cite{Abrahametal2016b}:
\begin{quote}
Extend the SMT-LIB language to cover a wider range of interests in the joint \scsc community. 
These of course include conjunctive arithmetic fragments on various (maybe mixed) domains. Among other potential extensions are: optimisation (finding a solution maximising a goal function), allowing the use of differential equation theory, simplification of formulas, quantifier elimination.
\end{quote}
Hence, at this stage in the development of SMT-LIB we consider what lessons can be learnt from the OpenMath community, and how both languages could be adapted to support the growing \scsc initiative.

\section{Background}

\subsection{MathML}

MathML is described in \cite{WorldWideWebConsortium2014a} and is usually encoded in XML.  A special aspect of MathML is that there are two main strains of markup: Presentation Markup is used to display mathematical expressions while Content Markup is used to convey mathematical meaning \cite[\S1.3]{WorldWideWebConsortium2014a}.

\subsection{OpenMath}

The OpenMathStandard is in \cite{Buswelletal2004}. A new version is in preparation, but the only substantive change is to clarify the relationship with MathML. OpenMath objects are seen as trees, and can be encoded in XML or in binary --- for readability we use the XML encoding here.  There are various basic objects: symbols \verb+OMS+, integers \verb+OMI+, 64-bit IEEE \cite{IEEE2008} floating point objects \verb+OMF+, uninterpreted byte arrays \verb+OMB+, strings \verb+OMSTR+ and variables \verb+OMV+, all of which are leaves of the tree; and various constructions: application \verb+OMA+, binding \verb+OMB+ and the statement of bound variables \verb+OMBVAR+, errors \verb+OME+, attributes \verb+OMATTR+ and attribution pairs \verb+OMATP+, and foreign objects \verb+OMFOREIGN+.
\par
The mathematical expression $x+1$ would be encoded as the application of the symbol '+' to $x$ and $1$:
\begin{verbatim}
<OMA>
  <OMS name="plus" cd="arith1"/>
  <OMV name="x"/>
  <OMI> 1 </OMI>
</OMA>
\end{verbatim}
As can be seen this is somewhat verbose. An alternative syntax, POPCORN \cite{HornRoozemond2009} has been proposed, which would shorten this algorithmically to \verb!arith1.plus($x,1)!, and then, knowing about \verb+arith1+, still further to \verb!$x+1!.

\subsubsection{Symbols}\label{OM:symbols}
We have written this explicitly with the integer 1. If, however, we wanted the multiplicative identity of whatever ambient algebra we were working in, we would have used the nullary symbol \verb+<OMS name="one" cd="alg1"/>+. \verb+alg1+ also defines \verb+zero+ as the additive identity.
\par
The symbol \verb+plus+ comes from the {\bf Content Dictionary} \verb+arith1+, which contains various basic mathematical operators, including a (not necessarily commutative) \verb+times+ operator. If one wants an explicitly commutative \verb+times+ operator, there is one in \verb+arith2+ which has the property that $x*y=y*x$, or, from the Content Dictionary, the {\bf Formal Mathematical Property} in Figure \ref{fig:FMP}. These Formal Mathematical Properties describe (some of) teh semantics of the OpenMath symbols.

\begin{figure}[ht]
\caption{Commutatiivity of {\tt times} from {\tt arith2}\label{fig:FMP}}
\begin{verbatim}
<FMP>
  <OMOBJ xmlns="http://www.openmath.org/OpenMath" version="2.0" 
         cdbase="http://www.openmath.org/cd">
    <OMBIND>
      <OMS cd="quant1" name="forall"/>
      <OMBVAR>
         <OMV name="a"/>
         <OMV name="b"/>
      </OMBVAR>
      <OMA>
        <OMS cd="relation1" name="eq"/>
        <OMA>
          <OMS cd="arith2" name="times"/>
          <OMV name="a"/>
          <OMV name="b"/>
        </OMA>
        <OMA>
          <OMS cd="arith2" name="times"/>
          <OMV name="b"/>
          <OMV name="a"/>
        </OMA>
      </OMA>
    </OMBIND>
  </OMOBJ>
</FMP>
\end{verbatim}
\end{figure}

\subsubsection{Small Type System}\label{sec:STS}

OpenMath \emph{per se} is type-agnostic. It is expected that serious type systems will build, parallel with the Content Dictionary system, a set of files describing the type system for the symbols.
There is a simple Small Type System described in \cite{Davenport2000c}. The entry for \verb+<OMS name="times" cd="arith2"/>+ (which comes from the file \verb+arith2.sts+) is given in Figure \ref{fig:STS}: it states that the symbol is nary and associative\footnote{Both properties have to be stated: set construction is nary but not associative, for example.}, takes arguments from a structure which has the property \verb+AbelianSemiGroup+, and returns an answer in the same \verb+AbelianSemiGroup+.

\begin{figure}[ht]
\caption{Small Type System for {\tt times} from {\tt arith2}\label{fig:STS}}
\begin{verbatim}
<Signature name="times">
<OMOBJ xmlns="http://www.openmath.org/OpenMath">
 <OMA>
  <OMS name="mapsto" cd="sts"/>
  <OMA>
   <OMS name="nassoc" cd="sts"/>
   <OMV name="AbelianSemiGroup"/>
  </OMA>
  <OMV name="AbelianSemiGroup"/>
 </OMA>
</OMOBJ>
</Signature>

\end{verbatim}
\end{figure}

\subsubsection{Binders}

OpenMath does not have a fixed set of binders. However, binders are introduced through a fixed syntactic marker \verb+OMBIND+ which enables correct recognition of free/bound variables. There is an example in Figure \ref{fig:FMP}, and further discussion in \cite{DavenportKohlhase2009c}.

\subsection{SMT-LIB}

This is described in \cite{Barrettetal2015b}, though a near-final draft of the next version is in \cite{Barrettetal2017a}. 

SMT-LIB specifies four languages:
\begin{enumerate}
\item a language for writing terms and formulas in a sorted (i.e., typed) version of first-order logic;
\item a language for specifying background theories and fixing a standard vocabulary of sort, function, and predicate symbols for them;
\item a language for specifying logics, suitably restricted classes of formulas to be checked for satisfiability with respect to a specific background theory;
\item a command language for interacting with SMT solvers via a textual interface that allows asserting and retracting formulas, querying about their satisfiability, examining their models or their unsatisfiability proofs, and so on.
\end{enumerate}
Of these, Language 1 corresponds to OpenMath, Languages 2 and 3, essentially, to a typing system as in Section \ref{sec:STS}, and Language 4 more to a command system such as SCSCP \cite{Lintonetal2013a}.

The syntactic encoding of SMT-LIB expressions (Language 1 above) is as LISP S-expressions\footnote{An argument for an XML encoding is made in \cite{MaricJanicic2004}, but there has been little progress on this recently.}:
\begin{verbatim}
<spec_constant> ::= <numeral> | <decimal> | <hexadecimal> | <binary> | <string>
<s_expr> ::= <spec_constant> | <symbol> | <keyword> | ( <s_expr>* )
\end{verbatim}
Just as in OpenMath, the mathematical expression $x+1$ would be encoded as the application of the symbol '+' to $x$ and $1$:
\begin{verbatim}
(+ x 1)
\end{verbatim}
where \verb#+# and \verb#1# would be defined in the logic of the file: the logic itself would refer to a theory for arithmetic.

\begin{quote}
Their [\verb+spec_constant+] semantics is determined locally by each SMT-LIB theory
that uses them. For instance, it is possible for an SMT-LIB theory of sets to use the numerals \verb+0+
and \verb+1+ to denote respectively the empty set and universal set. Similarly, the elements of \verb+binary+
may denote integers modulo n in one theory and binary strings in another; the elements of
\verb+decimal+ may denote rational numbers in one theory and floating point values in another.
\end{quote}
This contrasts with OpenMath's behaviour, as described in Section  \ref{OM:symbols}: OpenMath would use \verb+<OMS name="one" cd="alg1"/>+ to get the effect SMT-LIB gets from \verb+1+.

\subsubsection{Well-Sorted Terms}
The sort declarations in SMT-LIB state the types of \verb+symbol+ operators, so
that we can state that an SMT-LIB formula is well-sorted with respect to a given
sort declaration.  Besides the sorts used in the theories, it is possible to
declare other sorts, whose domain is uninterpreted, that is, they can be
interpreted as any non-empty set of elements.  Also, besides the functions and
predicates provided by the theories, uninterpreted functions and predicates can
be used; these can have uninterpreted sort either as a domain or as range, but
SMT-LIB makes it also possible to declare, for instance, a unary function from
integer to integer that is arbitrary.

\subsubsection{Binders}
SMT-LIB has precisely three binders: \verb+let+, \verb+forall+ and \verb+exists+. The last two are \emph{sorted}, in the sense that the syntax is 
$$
(\verb+forall+\ ((x_1\ \sigma_1) \cdots (x_n\ \sigma_n))\ \phi).
$$
The semantics are those of nested unary \verb+forall+, so that, while the $x_i$ may be repeated, earlier occurrences are shadowed by later ones. For \verb+let+, the syntax is 
$$
(\verb+let+\ ((x_1\ \tau_1) \cdots (x_n\ \tau_n))\ \phi),
$$
equivalent to the mathematical $\phi[\tau_1/x_1,\ldots,\tau_n/x_n]$ with simultaneous substitution. Hence the $x_i$ must be distinct in this case.

\section{Exists Uniquely}

A relatively recent\footnote{A textbook usage is \cite[p. 3]{Aluffi2009}, but it is more commonly found in research papers.} addition to mathematical notation is $\exists!$, meaning ``exists uniquely''. It is, of course, not logically necessary: two alternative definitions\footnote{There are more perverse but more compact ones.} are as follows:
\begin{equation}\label{eqA}
\exists! P(x) \Leftrightarrow \exists x \left(P(x)\land\forall y (P(y)\Rightarrow x=y)\right)
\end{equation}
\begin{equation}\label{eqB}
\exists! P(x) \Leftrightarrow \left(\exists x P(x)\right)\land\left(\forall y\forall z P(y)\land P(z)\Rightarrow y=z\right)
\end{equation}
Considered computationally, (\ref{eqA}) introduces an alternation, but fewer distinct quantifiers, and fewer repetitions of $P$ than (\ref{eqB}). 

\subsection{OpenMath}\label{sec:OMexists}

Since it is both useful and economical (saving the repetition of $P$, and the human/computer needing to recognise that it is the same $P$), there seems no reason not to introduce it.

\subsection{SMT-LIB}\label{sec:SMTexists}

Here the argument is more finely balanced. The arguments for are the same as for OpenMath (except that an SMT solver is expected to have clever heuristics, and idiom recognition might well be one of those).  The converse argument is that adding syntactic sugar is adding noise too.

\section{Maxima}
\subsection{OpenMath}\label{sec:OMmax}

The fundamental construct in OpenMath is \verb+minmax1.max+, which returns the maximum of a set. Hence we could encode $\max_{x\in[0,1]}x(1-x)$ (whose value is $\frac14$) as the following.
\begin{verbatim}
<OMA>
  <OMS cd="minmax1" name="max"/>
  <OMA>
    <OMS cd="set1" name="map"/>
    <OMBIND>
      <OMS cd="fns1" name="lambda"/>
      <OMBVAR>
        <OMV name="x"/>
      </OMBVAR>
      <OMA>
        <OMS cd="arith1" name="times"/>
        <OMV name="x"/>
        <OMA>
          <OMS cd="arith1" name="minus"/>
          <OMI> 1 </OMI>
          <OMV name="x"/>
        </OMA>
      </OMA>
    </OMBIND>
    <OMA>
      <OMS cd="interval1" name="interval_cc"/>
      <OMI> 0 </OMI>
      <OMI> 1 </OMI>
    </OMA>
  </OMA>
<OMA>
\end{verbatim}
This is a perfectly legitimate encoding, but one could argue that it is not very constructive, since the maximum is being taken over an uncountable set. Essentially, \verb+minmax1.max(set1.map(...))+ is an idiom for ``use the calculus operational semantics of $\max$'', except of course when it isn't.  It would be more helpful to have an explicit $\max$ constructor that took a set and a function. There are essentially two options here (OpenMath could, of course, adopt both).
\begin{enumerate}
\item An operator that took both a set and a function: essentially making explicit the idiom \verb+minmax1.max(set1.map(...))+ referred to above.
\item A binder that took both a function body and a predicate, using the same bound variable for both.
\begin{verbatim}
<OMBIND>
  <OMS cd="minmax2" name="max"/>
  <OMBVAR>
    <OMV name="x"/>
  </OMBVAR>
  <OMS cd="arith1" name="times"/>
  <OMV name="x"/>
  <OMA>
    <OMS cd="arith1" name="minus"/>
    <OMI> 1 </OMI>
    <OMV name="x"/>
  </OMA>
  <OMA>
    <OMS cd="set1" name="in"/>
    <OMV name="x"/>
    <OMA>
      <OMS cd="interval1" name="interval_cc"/>
      <OMI> 0 </OMI>
      <OMI> 1 </OMI>
    </OMA>
  </OMA>
<OMA>
\end{verbatim}
\end{enumerate}
The second one probably has the abvantage of being closer to common usage.
\par
A further complication is the lack of distinction between $\max$ and $\inf$.

\subsection{SMT-LIB}

SMT-LIB does not have a $\max$ operator.  However, OptiMathSAT's input language \cite{SebastianiTrentin2015a} is SMT-LIB extended with \verb+maximize+, \verb+minimize+ ``commands''.  In fact, these are statements as to the nature of the goal(s), and the goal is achieved by \verb+check-sat+.


\subsection{$\bm{\argmax}$}
\label{sec:argmax}

A relatively recent piece of mathematical notation is $\argmax$, which does not have an extremely formal definition. The Wikipedia\footnote{\url{https://en.wikipedia.org/wiki/Arg_max} [20th June 2017].} definition is that these ``are the points of the domain of some function at which the function values are maximized.'' Hence na\"\i{}vely,
\[
\displaystyle\argmax_{x\in\R}\sin(x)=\left\{\frac\pi2+2n\pi|n\in\Z\right\}.
\]
Though it can be defined in terms of other objects, it might be helpful to have a \verb+argmax+ constructor in OpenMath, capable of encoding $\argmax_{x\in[0,1]}x(1-x)$ (whose value is $\left\{\frac12\right\}$) .
\par
This is a perfectly sound mathematical definition, but does not really meet the requirements of \scsc, or computation in general. What \scsc really needs is a \emph{witness} point, i.e. a single value $x_0$ such that $f(x_0)=\max_{f\in S}f(x)$. For the sake of mathematical notation, we term this $\argmax^{(1)}$ --- one important point would be that it is not necessarily deterministic. This constructor could be called \verb+argmaxone+.

The OptiMathSAT approach is, \emph{after} calling \verb+check-sat+, to allow \verb+(get-value argument1)+ etc. to find the values at which the maximum discovered by \verb+check-sat+ was achieved.  This is essentially an $\argmax^{(1)}$ approach.  The precise details are more technical, as multiple maxima can be searched for: see Appendix \ref{Ex}.

\section{Output formats}
OpenMath is agnostic about whether its formulae are input or output: it is aimed at a world where one system's output is the next system's input. SMT-LIB, by contrast, is largelythe input language for an SMT system. The output is, curdely, either \verb+SAT+ or \verb+UNSAT+. However, \verb+SAT+ should have a model produced, i.e. a constructive demonstration of satisfiability.

The SMT-LIB standard does not specify what a model should be.  
There is thus the possibility to use any term, possibly to build algebraic numbers, etc...  However, the current SMT-LIB requires that two terms with different expressions in the description of the model have a different value in the model.  So it would be necessary to use some form of canonicalization of algebraic numbers that guarantees this.  


\section{Conclusions}

\subsection{Recommendations to OpenMath}

\begin{enumerate}
\item Formalise the r\^ole of POPCORN in the OpenMath stable.
\item Consider  an explicit $\exists!$ constructor (Section \ref{sec:OMexists}).
\item Consider  an explicit $\max$ constructor that took a set and a function (Section \ref{sec:OMmax}). OpenMath would then need to decide which variant(s) to adopt.
\item Clarify distinction between $\max$ and $\inf$ (Section \ref{sec:OMmax}).
\item Consider  an explicit $\argmax$ constructor (Section \ref{sec:argmax}).
\item Consider  an explicit $\argmax^{(1)}$ constructor (Section \ref{sec:argmax}).
\end{enumerate}

\subsection{Recommendations to SMT-LIB}

\begin{enumerate}
\item Consider having an operator to express quantifier elimination. This would also mean expressing the output, which could be done by reusing the input language\footnote{At least over the reals, by the Tarski--Seidenberg Theorem.}.
\item Consider  an explicit $\max$ constructor.
\item Consider  an explicit $\argmax^{(1)}$ constructor.
\end{enumerate}

\subsubsection*{Acknowledgements}

We are grateful to Pascal Fontaine (LORIA) for many useful comments.

We are grateful for support by the H2020-FETOPEN-2016-2017-CSA project
\scsc (712689).

\bibliographystyle{alpha} 
\bibliography{../../../../jhd}

\newpage

\appendix

\section{OptiMathSAT Example}\label{Ex}
\begin{verbatim}
(set-option :produce-models true)

(declare-fun cost () Real)
(declare-fun s1 () Bool)
(declare-fun s2 () Bool)
(declare-fun s3 () Bool)
(declare-fun s4 () Bool)
(declare-fun q1 () Real)
(declare-fun q2 () Real)
(declare-fun q3 () Real)
(declare-fun q4 () Real)

; set goods quantity
(assert (= 250 (+ q1 q2 q3 q4)))

; set goods offered by each supplier
(assert (or (= q1 0) (and (<= 50 q1) (<= q1 250))))
(assert (or (= q2 0) (and (<= 100 q2) (<= q2 150))))
(assert (or (= q3 0) (and (<= 100 q3) (<= q3 100))))
(assert (or (= q4 0) (and (<= 50 q4) (<= q4 100))))

; supplier is used if sends more than zero items
(assert (and (=> s1 (not (= q1 0))) (=> s2 (not (= q2 0)))
           (=> s3 (not (= q3 0))) (=> s4 (not (= q4 0)))))

; supply from the largest number of suppliers
(assert-soft s1 :id unused_suppliers)
(assert-soft s2 :id unused_suppliers)
(assert-soft s3 :id unused_suppliers)
(assert-soft s4 :id unused_suppliers)

; set goal (A)
(minimize (+ (* q1 23) (* q2 21) (* q3 20) (* q4 10)))
; set goal (B)
(minimize (+ (* q1 25) (* q2 19) (* q3 10) (* q4 20)))

; box: independent optimization
(set-option :opt.priority box)
(check-sat)

; print model for A
(set-model 0)  
(get-value (q1))
(get-value (q2))
(get-value (q3))
(get-value (q4))

; print model for B
(set-model 1)
(get-model)
\end{verbatim}
\end{document}